\documentclass[pre,reprint,twocolumn,superscriptaddress,nofootinbib]{revtex4-1} 

\usepackage{amsfonts,amsmath,amssymb}
\usepackage[]{graphicx}
\usepackage{floatrow}
\usepackage{textcomp,verbatim,color,wasysym,MnSymbol}
\usepackage[english]{babel}
\usepackage{siunitx}
\usepackage{enumitem}
\usepackage{pifont}
\usepackage{placeins}
\usepackage{soul}

\newcommand{\VECf}{{\boldsymbol{f}}}

\newcommand{\VECn}{{\boldsymbol{n}}}
\newcommand{\VECt}{{\boldsymbol{t}}}
\newcommand{\VECu}{{\boldsymbol{u}}}
\newcommand{\VECx}{{\boldsymbol{x}}}

\newcommand{\VECnab}{{\boldsymbol{\nabla}}}

\newcommand{\CALC}{{\mathcal{C}}}
\newcommand{\CALG}{{\mathcal{G}}}

\newcommand{\NN}{{\mathbb{N}}}

\begin{document}

\title{Vesicle dynamics in confined steady and harmonically modulated Poiseuille flows}

\author{Zakaria Boujja}
\email{boujja.zakaria@gmail.com}
\affiliation{Experimental Physics, Saarland University, Saarbr\"{u}cken, Germany.}
\affiliation{LaMCScI, Universit\'e Mohamed V, Facult\'e des Sciences, Rabat, Morocco.}
\author{Chaouqi Misbah}
\affiliation{LIPHY, Universit\'e Grenoble Alpes, F-38000 Grenoble, France.}
\author{Hamid Ez-Zahraouy}
\affiliation{LaMCScI, Universit\'e Mohamed V, Facult\'e des Sciences, Rabat, Morocco.}
\author{Abdelilah Benyoussef}
\affiliation{LaMCScI, Universit\'e Mohamed V, Facult\'e des Sciences, Rabat, Morocco.}
\author{Thomas John}
\affiliation{Experimental Physics, Saarland University, Saarbr\"{u}cken, Germany.}
\author{Christian Wagner}
\affiliation{Experimental Physics, Saarland University, Saarbr\"{u}cken, Germany.}
\affiliation{Physics and Materials Science Research Unit, University of Luxembourg, Luxembourg.}
\author{Martin Michael M\"{u}ller}
\affiliation{Laboratoire de Physique et Chimie Th\'eoriques - UMR 7019, Universit\'e de Lorraine, 1 boulevard Arago, F-57070 Metz, France.}
\date{\today }


\begin{abstract}
We present a numerical study of the time-dependent motion of a two-dimensional vesicle in a channel under an imposed flow. In a Poiseuille flow the shape of the vesicle depends on the flow strength, the mechanical properties of the membrane, and the width of the channel as reported in the past. This study is focused on the \emph{centered snaking (CSn)} shape, where the vesicle shows an oscillatory motion like a swimmer flagella even though the flow is stationary. We quantify this behavior by the amplitude and frequency of the oscillations of the vesicle's center of mass. We observe regions in parameter space, where the CSn coexists with the parachute or the unconfined slipper. The influence of an amplitude modulation of the imposed flow on the dynamics and shape of the snaking vesicle is also investigated. For large modulation amplitudes transitions to static shapes are observed. A smaller modulation amplitude induces a modulation in amplitude \textit{and} frequency of the center of mass of the snaking vesicle. In a certain parameter range we find that the center of mass oscillates with a constant envelope indicating the presence of at least two stable states.

\end{abstract}

\maketitle


\section{Introduction}

Blood is a viscous, non-homogeneous, and complex fluid \cite{baskurt2007handbook}. The most important components are red blood cells (RBC), or erythrocytes. It is thus not surprising that 
their dynamics plays an important role in blood flow in general \cite{koeppen1992berne,openstax2017anatomy}. Normal human RBCs in the undeformed state 
are elastic capsules of biconcave disk-like shape whose maximum diameter is of the order of ten microns \cite{barthes2016motion}.  
The surface area to volume ratio of the normal cell is 40\% greater than that of a sphere with the same volume \cite{eggleton1998large}, 
which not only improves the RBCs efficiency in loading and unloading solutes, but also allows the cells to deform 
easily. This is of eminent importance as they must pass the circulatory system where they encounter capillaries with diameters as small as 2.5 $\mu$m \cite{baskurt2007handbook,openstax2017anatomy}.
The interior cell fluid, called the cytoplasm, is enclosed by a fluid bilayer membrane made of phospholipids with a supportive cytoskeleton of proteins underneath. The whole envelope is nearly incompressible and resists shear which allows the RBC to recover its initial biconcave 
shape after a deformation by external forces. 

Simplified systems, like vesicles made of a pure bilayer of phospholipids and capsules made of an extensible polymer shell, are used as models for RBCs. Both systems reproduce several features of RBCs remarkably well \cite{barthes2016motion,pozri2003numerical,misbah2012vesicles,guckenberger2017theory}. In a steady linear shear flow, for example, one observes three types of motion: $(i)$ a tank-treading motion in which the cell membrane and the interior liquid rotate, while the cell aligns at an angle with the flow direction, $(ii)$ a tumbling motion that is characterized by the flipping of the cell resembling a rigid-body motion, and $(iii)$ an intermediate regime termed vacillating-breathing (also called swinging or trembling mode), in which the long axis oscillates around the flow direction, while the shape undergoes a
breathing-like motion \cite{fischershear,abkarianshear}.
Under a Poiseuille flow, a case that is relevant for blood circulation, RBCs assume symmetric or asymmetric static shapes, in particular the so-called parachute and slipper shapes. One also observes oscillatory shapes such as the snaking shape \cite{abkarian2008poseuille,vlahovska2009vesicles}. 
In order to reproduce these shapes in theoretical studies, it is necessary to include the interactions between the fluid and the bilayer membrane. In the past, several analytical and numerical methods have been developed. In all calculations the Stokes approximation for flow at low Reynolds number was used. Most analytical approaches are limited to situations where the shape is close to a sphere. This has been successfully employed in studies of single vesicles \cite{misbah2006vacillating,vlahovska2007dynamics},
and capsules \cite{barthes1981time,finken2011micro} under unbounded shear and Poiseuille flows. 
From a numerical point of view, boundary integral methods (BIM) are one of the most powerful and effective techniques to solve flow problems with arbitrary boundary conditions \cite{pozrikidis1992boundary}. They are based on the use of Green's functions and have been applied  
for vesicles in 2D \cite{kaoui2011complexity,veerapaneni2009boundary} and 3D \cite{biben2011three,zhao2011dynamics, boedec20113d} as well as for capsules \cite{breyiannis2000simple,lac2004spherical,guckenberger2017numerical}. Another method combines a particle-based hydrodynamics model for the fluid with a coarse-grained surface model for the membrane \cite{Noguchi2005,Noguchi2011}. 
Alternatively, one can mesh the fluid domain using finite elements or a lattice Boltzmann method. The cell is then immersed in that domain \cite{peskin1977numerical,eggleton1998large,sui2007transient}. The associated method is called the immersed boundary method. 

In this study we consider a two-dimensional vesicle in a channel \cite{aouane2014,aouane2017}. To include the elastic rigidity of the vesicle, we use the Helfrich model \cite{Helfrich1973} to account for bending together with a constraint expressing the inextensibility of the membrane. The flow field, the forces on the membrane and the resulting time-dependent deformations are solved with the generalized BIM \cite{pozrikidis1992boundary} as described in Sec.~\ref{sec:themodel}.
This model predicts the parachute, slipper and the snaking shape as a function of the flow strength, the bending rigidity, and the channel width in steady Poiseuille flow \cite{kaoui2011complexity,tahiri2013,aouane2014,guckenberger2017numerical}.
In the following we focus on the centered snaking (CSn) shape. Previous studies have already observed this shape for vesicles \cite{kaoui2011complexity,tahiri2013,aouane2014} and for capsules \cite{fedosov2014deformation}. However, they did not study it in detail, which motivates our study. 
In Sec.~\ref{sec:steadyPoiseuille} we will present and discuss the oscillation of the CSn in steady Poiseuille flow as a function of two characteristic parameters, the capillary number and the confinement. In addition to what was already known \cite{tahiri2013,aouane2014} we find domains of coexistence between the CSn and the parachute or the unconfined slipper. 
In Sec.~\ref{sec:modulatedPoiseuille} we will extend the discussion to time-dependent flows using an amplitude modulation of the imposed Poiseuille flow and investigating this effect on the vesicle's shape. 


\section{The model\label{sec:themodel}}
At the scales and velocities of blood flow in the capillaries of organisms viscous forces are dominant over the effects of inertia \cite{Skalak1989}. The Reynolds number is much smaller than one, which is why the fluid inside and outside the membrane vesicle can be described by the (quasi\nobreakdash-)steady Stokes equations
\begin{subequations}\label{eq:Stokes}
        \begin{eqnarray}
    &- \VECnab P + \eta \Delta \VECu = \VECf \; ,&
  \\
    &\VECnab \cdot \VECu = 0 \; ,&
  \end{eqnarray}
  \end{subequations}
where $\VECu$ is the velocity field, $P$ the pressure field, $\eta$ the viscosity of the fluid, and $\VECf$ the membrane force density which is 
only non-zero on the contour of the vesicle. 

The elastic energy of the vesicle is at the origin of the membrane force density. It is mainly stored in the bending modes of the membrane. 
Together with the assumption that the membrane is inextensible, the 2D form of the corresponding energy functional can be written as \cite{Helfrich1973,Evans1974}
  \begin{equation}\label{eq:Helfrich}
    E=\frac{\kappa}{2} \oint_{\CALC} c^{2} ds + \oint_{\CALC} \zeta ds
    \; .
  \end{equation}
The two integrals run over the contour of the membrane $\CALC$. The first integral is a quadratic functional of the local curvature $c$ 
where $\kappa$ denotes the bending rigidity. The inextensibility constraint is imposed in the second integral via a local Lagrange multiplier $\zeta$. 

A functional derivative of Eq.~\eqref{eq:Helfrich} leads to the 2D membrane force density \cite{Kaoui2008,Guven2008}:
\begin{equation}\label{eq:Helfrich-f}
    \VECf= \left [\kappa \left(\frac{\partial^{2} c}{\partial s^{2}} + \frac{c^{3}}{2}\right)-c\, \zeta \right ] 
       \textbf{n} + \frac{\partial \zeta}{\partial s} \, \textbf {t}  \; ,
\end{equation}
where $\VECn$ is the normal vector pointing towards the exterior of the vesicle, and $\VECt$ the unit tangent vector pointing in the direction of increasing arc length.

The perimeter $p$ of the vesicle is fixed due to the inextensibility of the membrane. This is also the case for the enclosed area $A$ since the fluid inside the vesicle is assumed to be incompressible. The reduced area $\nu=A/\left(\pi \left(\frac{p}{2 \pi}\right)^{2}\right)$,
which is the ratio of the actual area enclosed by the vesicle over that of a disk having the same perimeter as the vesicle, is thus a constant as well.
In the following we will set $\nu$ to 0.6, similar to the values found for RBC \cite{Canham1970}.

We consider a membrane vesicle in a channel of width $W$ under an imposed (external) flow $\VECu^{\infty}$ 
whose Cartesian components are given by:
\begin{subequations}\label{eq:pois}
   \begin{eqnarray}
                u_{x}^{\infty} &=& u_\text{max} \left(1 - 4 \cfrac{y^{2}}{W^{2}}\right) [1+\varepsilon_{m} \cos(2\pi f_{m} t)] \; ,\\
                u_{y}^{\infty} &=& 0 \; ,
    \end{eqnarray} 
\end{subequations}
where $u_\text{max}$ is the maximal velocity occurring in the middle of the channel. The parameters $\varepsilon_{m}$ and $f_{m}$, respectively, represent the amplitude and the frequency of a harmonically modulated Poiseuille flow. 
This type of oscillating flow was already used to study capsule deformation in shear flow \cite{matsunaga2015deformation}. Noguchi et al.\  mimicked the oscillating Poiseuille flow using a microchannel whose width varies periodically \cite{noguchi2010dynamics}. For $\varepsilon_{m}=0$ we recover the usual steady Poiseuille flow. 

In all our simulations the viscosities of the fluid inside and outside of the vesicle are equal. 
\textit{In vivo} the inner viscosity for RBCs is around five time larger than the outer one. A viscosity ratio equal to one results in the same shapes as \textit{in vivo} in addition to coexistence regions (see below and Figs.~2 and 3 of \cite{tahiri2013}). 
As the aim of our study is to focus on the shapes and their dynamics we have chosen to neglect the effect of a viscosity difference. However, we expect our results to be valid at least qualitatively for the \textit{in vivo} value as well.

We only consider frequencies $f_m$ which are much smaller than the characteristic frequency of the flow field in the channel. This implies that we can assume a quasi-steady Stokes flow. In order to solve the corresponding Stokes equations, Eqs.~\eqref{eq:Stokes}, for this case we use the boundary integral method (BIM), based on the use of Green's functions \cite{Pozrikidis1992,matsunaga2015deformation}. The advantage of this method is that there is no need to solve for the fluid domain, and the whole dynamics of the vesicle is encoded in the membrane itself.
In this work we use a special Green function that automatically satisfies the no-slip boundary conditions at the boundaries for 2D Stokes flow \cite{blake1971note,Pozrikidis1992,liron1976stokes,Thiebaud2013}. 
The velocity field of a point $\VECx$ that may lie in the fluid or at the membrane is then given by:
         \begin{equation}\label{eq:BIM}
        \VECu (\VECx) = \VECu^{\infty}(\VECx)+
           \frac{1}{4\pi\eta} \oint_{\CALC} \CALG^{2w}(\VECx_{0},\VECx) \VECf(\VECx_{0}) \mathrm{d}s (\VECx_{0})
           \; ,
          \end{equation}  
where $\VECx$ and $\VECx_{0}$ are 2D position vectors and $\CALG^{2w}$ is the Green's function for a domain confined between two parallel planar walls, and the integral runs over the membrane contour $\CALC$ 
\footnote{More details on the numerical method can, for instance, be found in Ref.~\cite{aouane2017}.}.

The behavior of the vesicle in the channel can be described with the help of two dimensionless numbers: 
$(i)$ the degree of confinement      
\begin{equation}\label{eq:Cn}
  C_{n}=\frac{2R_{0}}{W} 
  \; ,
\end{equation}
and
$(ii)$ the capillary number, which measures the flow strength over the bending energy of the membrane      
\begin{equation}\label{eq:Ck}
  C_{k}=\frac{\eta R_{0}^{3} \dot{\gamma}}{\kappa} = \dot{\gamma} \tau_{\kappa}
  \; ,
\end{equation}    
where $\tau_{\kappa}$ is the time scale of the relaxation of the bending modes, which will be used to scale frequencies and simulation time.

To scale the other quantities of our system, we use $R_{0}=\sqrt{A/\pi}$ as the characteristic length scale and the local 
shear rate of the applied flow, $1 / \dot{\gamma}$, as the time scale of the flow. 
Note that we refrain from introducing an additional notation for scaled variables in the following to avoid overloading the notation.


\section{Results and Discussion}

\subsection{\emph{Steady Poiseuille flow}\label{sec:steadyPoiseuille}}

\begin{figure}
  \includegraphics[width=\columnwidth]{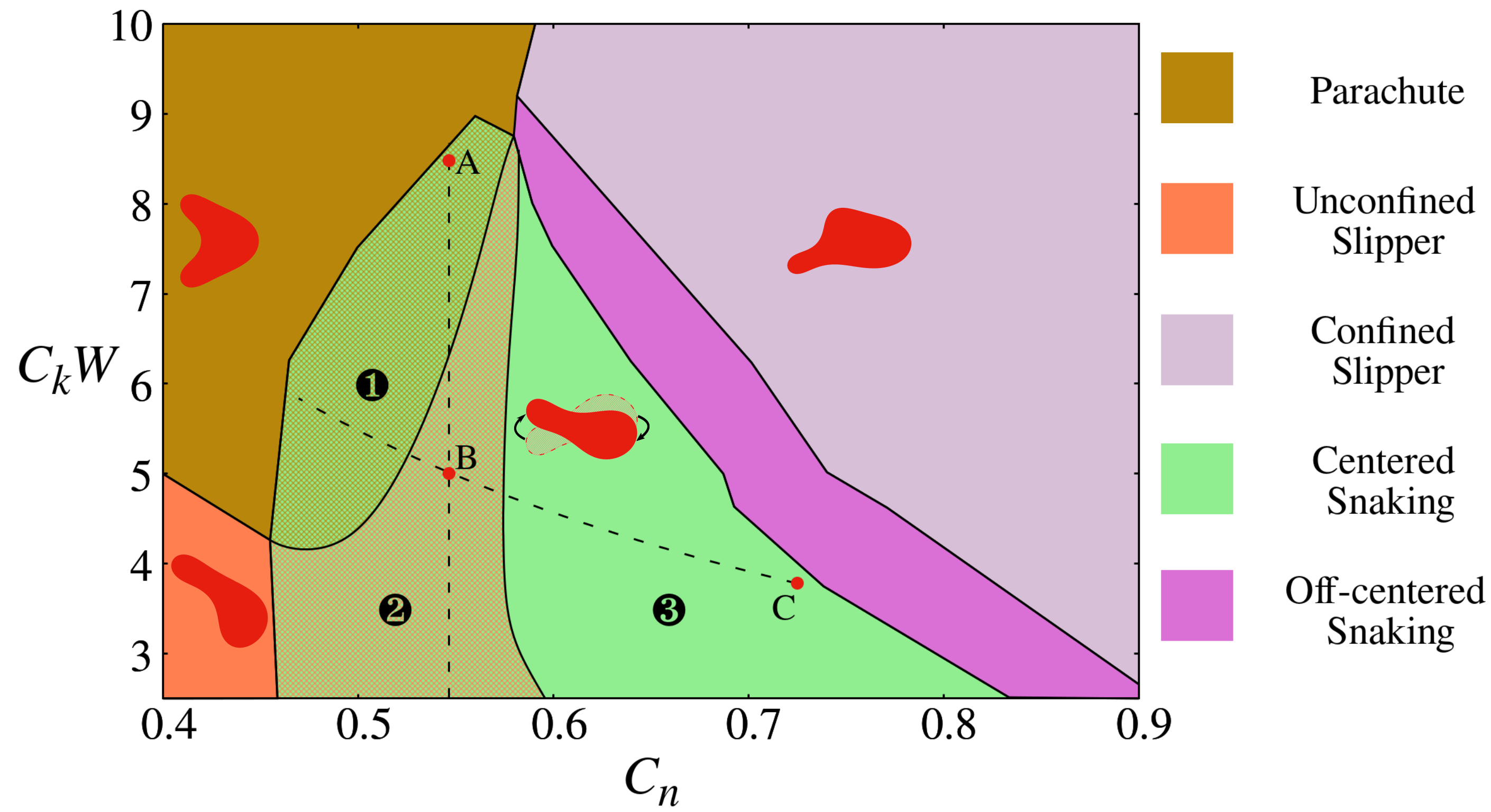}
  \caption{Phase diagram showing the different shapes of a vesicle in a steady Poiseuille flow as a function of $C_{k}W$ and $C_{n}$ (compare Refs.~\cite{tahiri2013} and \cite{aouane2014}). Note that $W$ is scaled by $R_0$ as explained in Sec.~\ref{sec:themodel}. The figure reproduces the regions of Figure 9(b) of Ref.~\cite{aouane2014} not showing the details of the off-centered snaking region. 
In addition to our earlier studies we find that the region containing the centered snaking shapes (CSn) consists of three different parts: two regions of coexistence (\ding{182}: CSn and parachutes, \ding{183}: CSn and unconfined slippers) and the region \ding{184}, where only CSn are found.}\label{phase_diag}
\end{figure}

The shapes of a single 2D vesicle in a channel under a steady Poseuille flow ($\varepsilon_{m}=0$ in Eqs.~(\ref{eq:pois})) have already been studied with the BIM to some detail in the past \cite{tahiri2013,aouane2014}. 
Depending on the capillary number $C_k$ and the confinement $C_n$ the vesicle either adopts a static shape or shows an oscillatory behavior as a function of time. 
The resulting shapes can be arranged in a phase diagram as was already shown in Ref.~\cite{aouane2014,tahiri2013}. In Fig.~\ref{phase_diag} we have reproduced the same phase diagram carefully checking the boundaries of each region. To scrutinize the CSn region in particular, we have performed simulations with different initial shapes and found coexistence regions between CSn and static shapes.
The static shapes are either symmetric or asymmetric. The center of mass of the former, like the parachute shape, always remains on the symmetry axis of the channel, while the latter, like the confined and unconfined slippers, have a center of mass which is displaced vertically. Additionally, one observes a tank-treading motion for the asymmetric shapes: their membrane is continuously rotating around the $z$ axis. Kaoui et al. have shown in a similar system that this tank-treading is due to the difference in velocity between the vesicle and the imposed flow \cite{kaoui2009}. 
A tangential motion of the membrane can also be observed for the oscillatory shapes. The membrane of the CSn, for example, displays a purely oscillatory tank-treading.   
The origin of this oscillation can be understood by having a closer look at the phase diagram (Fig.~\ref{phase_diag}). The region of the CSn is located at weak external flows ($C_{k}W<9$). When the walls of the confinement are sufficiently far away from the vesicle ($C_n\lesssim 0.45$), one either observes stable parachutes or unconfined slippers. Decreasing the size of the channel enhances the interaction between the channel and the vesicle. 
During the simulation the vesicle evolves towards a stable shape, but is repelled by the fluid interactions with one wall towards the other one. It again tries to adopt a stable shape but is pushed back again. The resulting oscillation gives rise to the CSn. When the channel is not yet too small ($0.45\lesssim C_n\lesssim 0.6$), 
the vesicle will not necessarily adopt the CSn but can become a parachute or an unconfined slipper as well, depending on the initial conditions of the simulation. 
The resulting regions of coexistence between the CSn and the parachutes/unconfined slippers are depicted in Fig.~\ref{phase_diag} \footnote{Note that regions of coexistence between parachutes and slippers have already been observed for vesicles whose inner fluid viscosity was five times the viscosity of the surrounding fluid \cite{tahiri2013}. We also searched our system for regions of coexistence in the parachute and both slipper regions but did not find any.}. For larger confinement ($C_n\gtrsim 0.6$), the coexistence ceases since the walls are so close that the unconfined shapes are not possible any more.

\begin{figure}
  \includegraphics[width=\columnwidth]{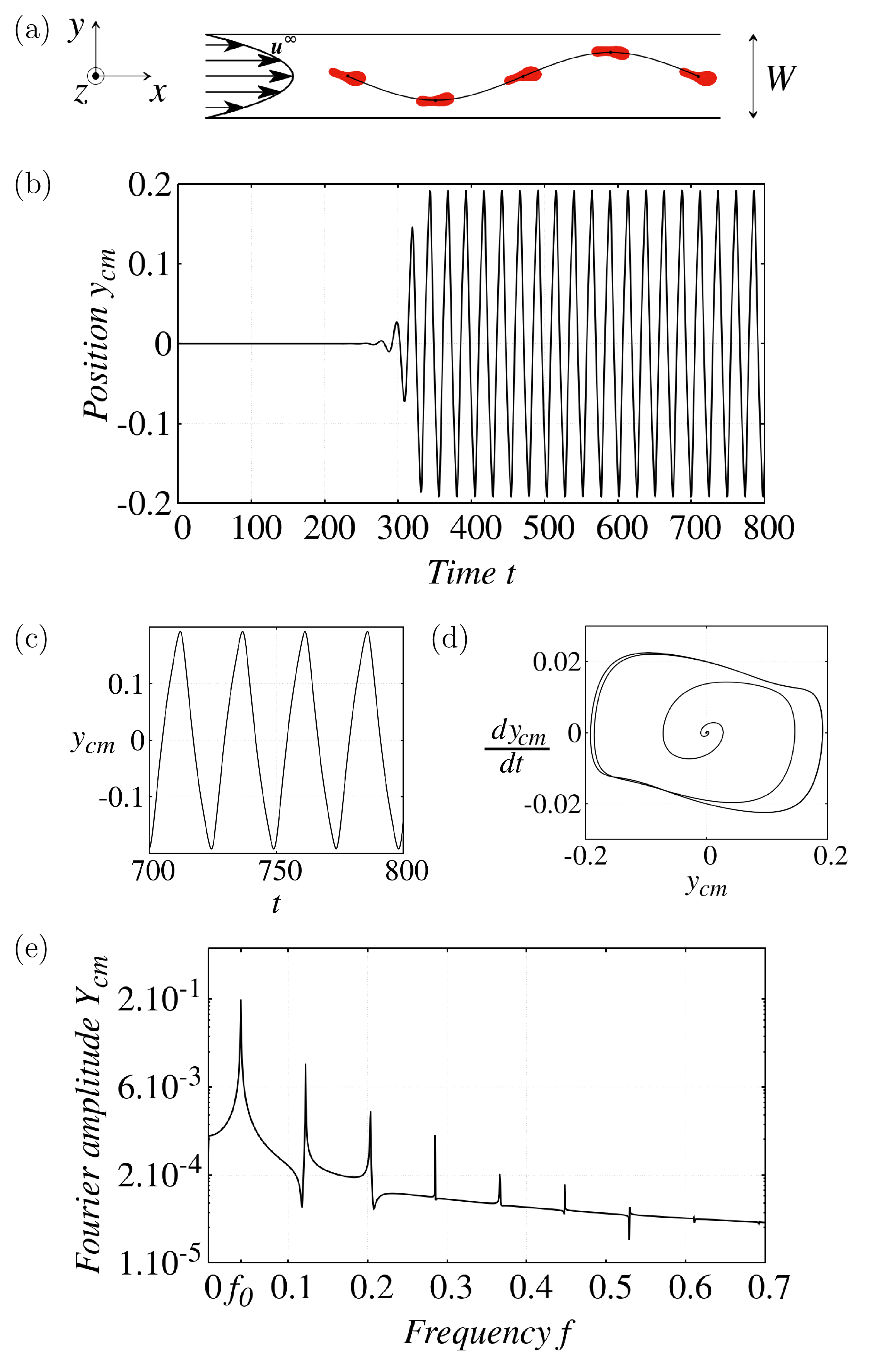}
  \caption{Centered snaking shape in a steady Poiseuille flow for $C_{k}W=5$ and $C_{n}=0.55$ ($cf.$ point B in Fig.~\ref{phase_diag}). (a) Sketch of the system. 
  (b) Vertical component of the vesicle's center of mass ($y_{cm}$) as a function of time. (c) Zoom of Fig.~\ref{Model}(b). (d) Poincar\'{e} map and 
  (e) Fourier transform of the signal of Fig~\ref{Model}(b). All variables in Figs. ~\ref{Model}(b) - ~\ref{Model}(e) are scaled as described in Sec.~\ref{sec:themodel}.
  } \label{Model}
\end{figure}

To characterize the oscillation of the CSn in more detail, we choose to look at the vertical position of the center of mass of the vesicle, $y_{cm}$, as a function of time. The maximal vertical distance of the center of mass from the mid-line, $y_{cm}^\text{max}$, is a measure for the amplitude of the CSn. 
Fig.~\ref{Model} shows an example for $C_{k}W=5$ and $C_{n}=0.55$ which corresponds to point B in Fig.~\ref{phase_diag}. After the transient
oscillation the vesicle starts oscillating periodically as one can see in Fig.~\ref{Model}(b) and the zoom Fig.~\ref{Model}(c). The Poincar\'{e}-map in Fig.~\ref{Model}(d) shows a non circular limit cycle, which implies that the oscillation of the vesicle is not harmonic. The Fourier transform of $y_{cm}$ in Fig.~\ref{Model}(c) confirms this observation. 

\begin{figure}
  \includegraphics[width=\columnwidth]{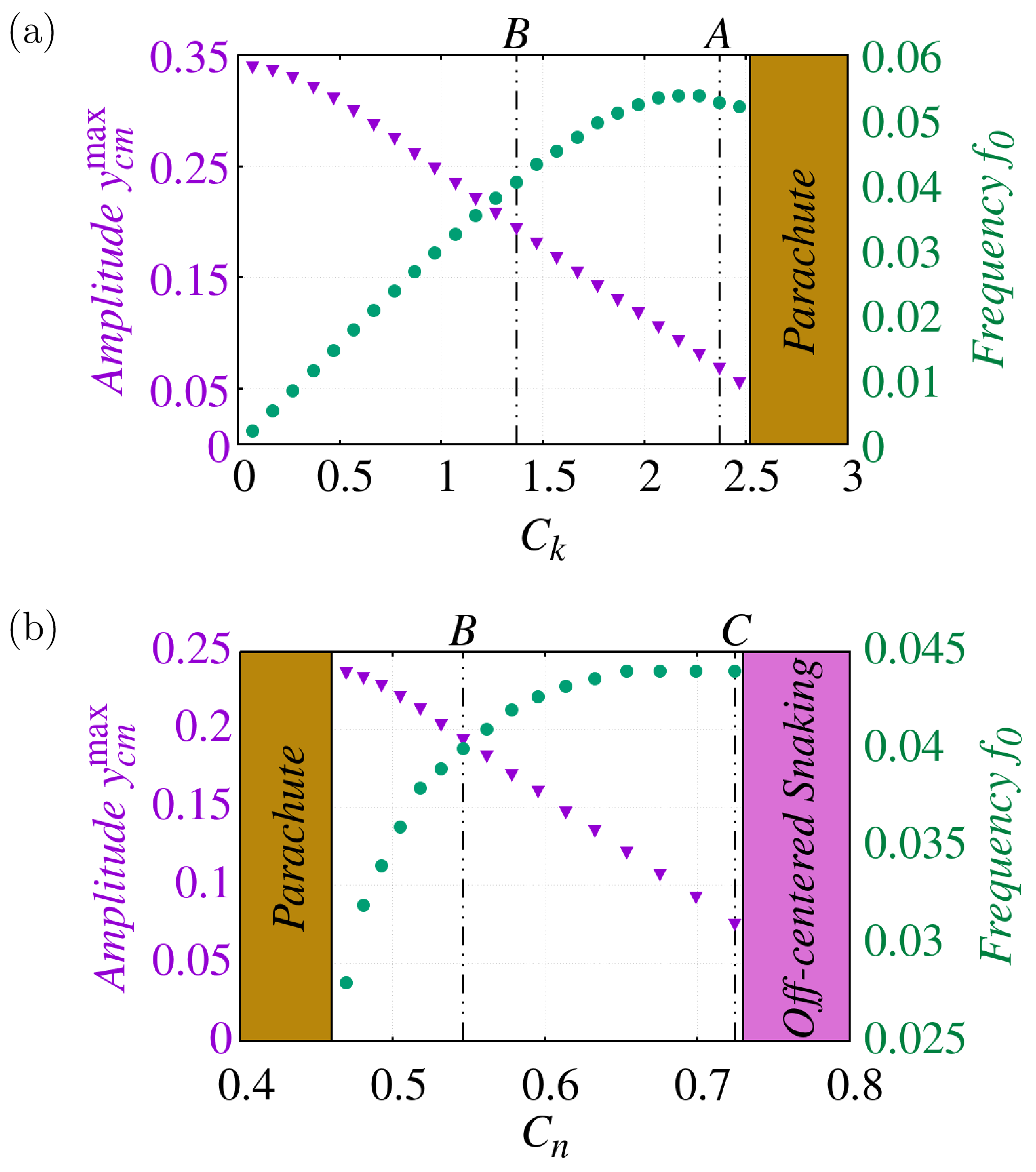}
  \caption{Scaled amplitude (triangles) and frequency (circles) of the oscillation of the CSn (a) for a fixed confinement $C_{n}=0.55$ and (b) for a fixed capillary number $C_{k}=1.37$. As an orientation you can find the points A, B, and C in the phase diagram, Fig.~\ref{phase_diag}, together with the two dashed curves corresponding to  $C_{n}=0.55$ and $C_{k}=1.37$, respectively. }\label{Ck_Cn_fix}
\end{figure}

An increase of $C_{k}$ or $C_{n}$ intensifies the interaction between the fluid and the membrane. The amplitude of the oscillation of the CSn will thus be attenuated as one can see in Fig.~\ref{Ck_Cn_fix}. At the same time we observe an increase of the fundamental frequency of the oscillation which becomes almost a constant close to the boundaries of the region of CSn.  In Fig.~\ref{Ck_Cn_fix}(a) the confinement is set to $C_{n}=0.55$ while $C_{k}$ is increased until the vesicle changes its shape to a parachute. In Fig.~\ref{Ck_Cn_fix}(b) we fix $C_{k}=1.37$ and increase $C_{n}$ (decreasing the channel width) 
until the vesicle reaches the off-centered snaking shape \textit{via} a complex dynamics (see Ref.~\cite{aouane2014} for the details).


\subsection{\emph{Harmonically modulated Poiseuille flow}\label{sec:modulatedPoiseuille}}

Red blood cells in circulation are subject to an unsteady flow. The intermittent nature of heart pumping causes the flow to pulsate. The unsteadiness can also arise due to the contraction and recoil of smaller arteries regulating local circulation \cite{koeppen1992berne} or diseased arteries with various degrees of atherosclerosis \cite{ku1985pulsatile}. 
There have been a few studies that address the effect of pulsatile flow on RBC dynamics \cite{silver1989estimation, mandelbaum1965pulsatile, ariman1974steady}. Nakajima et al., for example, have studied the deformation response of RBCs in a sinusoidally varying shear flow generated in a cone-and-plate viscometer \cite{nakajima1990}. 
A major finding of their experiment is that the deformation response is not identical during different phases of the shear flow. The deformation is higher during the retarding phase and lower during the accelerating phase. They noted that such an unequal response was probably due to the rheological properties of the intracellular fluid and its interaction with the membrane.

In the present article we will mimick the time dependence of the blood flow using Eqs.~(\ref{eq:pois}). The same type of flow has been used by A. Farutin and C. Mibah for an analytical study of the rheological properties of a single vesicle in shear flow \cite{farutin2012}. They have shown theoretically that the effective viscosity exhibits a resonance for vesicles, similarly to what happens for capsules \cite{kessler2009}. The amplitude $\varepsilon_{m}$ of the flow is now nonzero. The effect of this parameter and the frequency $f_{m}$ will be studied in the following.

To this end we have chosen three points in the phase diagram (see Fig.~\ref{phase_diag}). Each point lies in one of the three regions where the centered snaking shape can be found. The first point, A, with $C_{k}=2.33$ and $C_{n}=0.55$ lies in the region of coexistence between the CSn and the parachutes close to the upper boundary of the region. The second point, B,  corresponds to the same confinement as point A but smaller capillary number $C_{k}=1.37$. It lies in the region of coexistence between the CSn and the unconfined slippers. Finally, we choose point C as a representative of the region where only CSn are found by keeping the capillary number of point B but increasing the confinement to $C_{n}=0.72$.

\subsubsection{\emph{Influence of the amplitude}\label{subsubsec:amplitude}} 

After the vesicle has reached the CSn in Poiseuille flow, we switch on the harmonically modulated flow using the profile of Eqs.~(\ref{eq:pois}). The frequency $f_{m}$ is fixed to the fundamental frequency in steady Poiseuille flow which can be read off from Fig.~\ref{Ck_Cn_fix}. The vesicles corresponding to the  three considered points are then perturbed by varying the amplitude $\varepsilon_{m}$. 
As the point A is near to the boundary of the parachute region, a small amplitude $\varepsilon_{m}=0.2$ with $f_{m}=0.052$ can already induce a migration to the parachute shape. When we impose a steady Poiseuille flow again, the vesicle stays in the parachute shape as expected since A lies in the region of coexistence between the two shapes.
The CSn of point B can as well be forced to evolve to the parachute shape but a much higher amplitude is needed ($\varepsilon_{m}$=1). 
After imposing a steady Poisseuille flow again the vesicle takes on the shape of an unconfined slipper since the point B lies in the region of coexistence between the CSn and the unconfined slippers. 
In point C the vesicle is more confined than in A and B. Therefore, we cannot force it away from the CSn even if we apply the same high perturbation ($\varepsilon_{m}=1$). 
The shapes found in the three points for steady Poiseuille flow after switching off the time-dependence of the flow are all in agreement with the dashed areas in Fig.~\ref{phase_diag}.

\subsubsection{\emph{Influence of the frequency}\label{subsubsec:frequency}} 
\begin{figure}
  \includegraphics[width=\columnwidth]{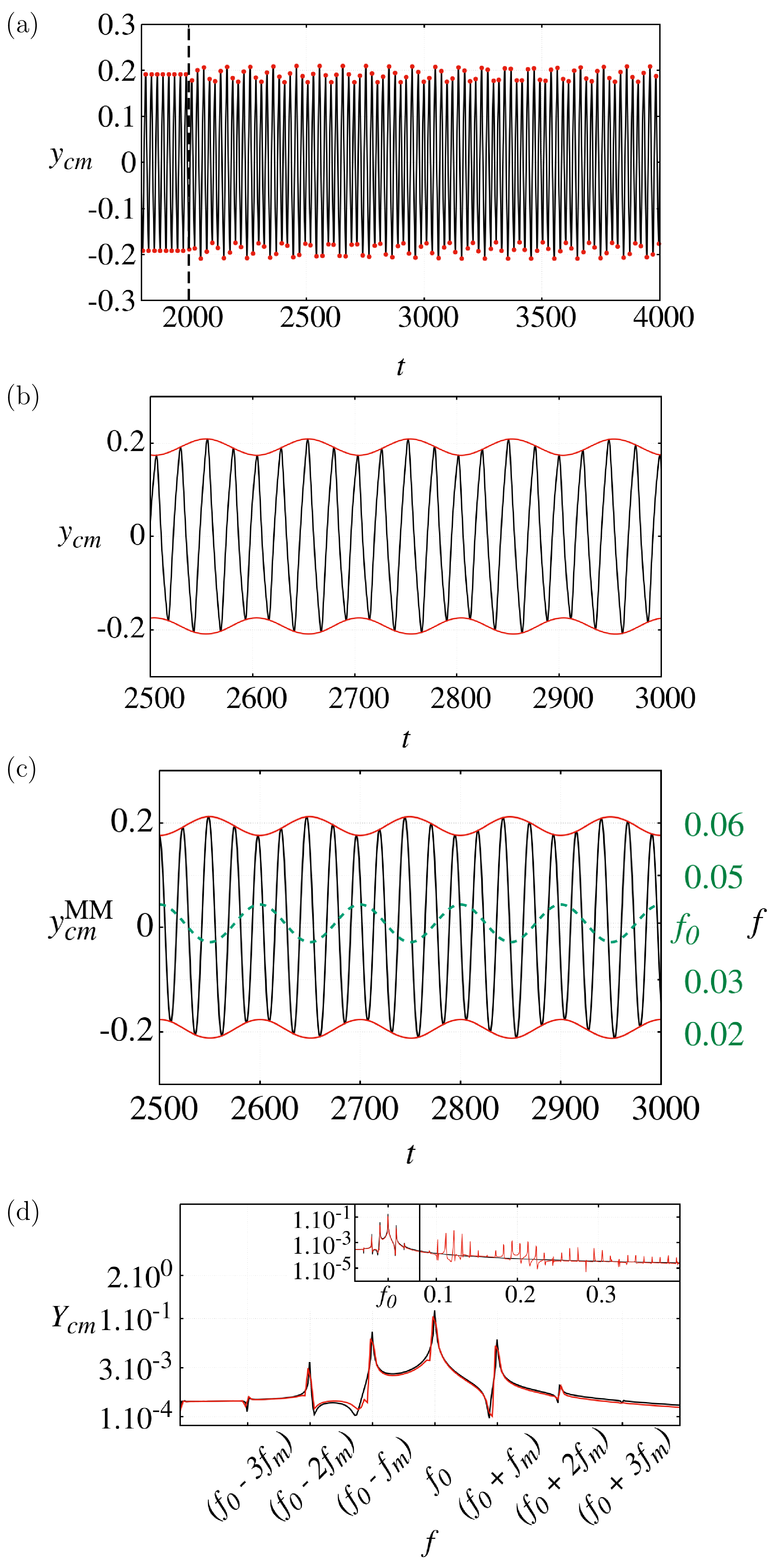}
  \caption{(a) Vertical component of the vesicle's center of mass from the simulation as a function of time in a harmonically modulated Poiseuille flow of amplitude  $\varepsilon_m=0.1$ and frequency $f_m=0.01$ around the point B. For this case $C_k^0 = 1.37$, $y_{cm}^0=0.19$, and $f_0 = 0.04$ (compare Figs.~\ref{phase_diag} and \ref{Ck_Cn_fix}(a)). The dashed line represents the time at which the modulation of the Poiseuille flow is switched on. 
(b) Zoom of Fig. ~\ref{MM}(a). (c) Corresponding mixed modulation signal $y_{cm}^{\text{MM}}$ (according to Eq.~(\ref{eq:mixedmodulation})) with indices of modulation $m=0.09$ and $\beta=0.4$ together with envelope (solid red) and frequency (dashed green). (d) Comparison of the Fourier transforms of theory (black) and simulation (red). All variables are scaled as described in Sec.~\ref{sec:themodel}.}\label{MM}
\end{figure}

To study the influence of the frequency, a similar procedure as described in the previous section is used in the simulations. We start with a CSn with a center of mass that oscillates periodically with amplitude $y^0_{cm}$ and fundamental frequency $f_0$ as described in Sec.~\ref{sec:steadyPoiseuille}. When it is put into a harmonically modulated flow of fixed amplitude $\varepsilon_{m}$ and frequency $f_m$, the oscillation becomes more complicated. 
Fig.~\ref{MM} shows one example for the point B with $\varepsilon_m=0.1$ and  $f_m = 0.01$. The investigated system is obviously not a linear time invariant (LTI) system, because the  
output contains non vanishing amplitudes at frequencies different from the input frequency \cite{DPSGuide,John2016}. 
However, at a first approximation, the nonlinear response of the system to the harmonically modulated flow can be treated using a quasi-stationary approach. The scaled frequencies $f_m$ and $f_0$ are both much smaller than one which implies that the elastic modes of the vesicle relax very quickly compared to a typical period of the shape oscillation. 
We can thus use the results of Sec.~\ref{sec:steadyPoiseuille} for a vesicle in steady Poiseuille flow. 
The time-dependent flow amplitude instantaneously affects the oscillation of the vesicle. 
Since the flow is directly proportional to the capillary number, we can use Fig.~\ref{Ck_Cn_fix}(a) to predict the behavior of the system. The imposed  oscillation of $C_k$ around the value of the initial CSn, $C_k^0$, 
enforces an oscillation of position and frequency of the vesicle's center of mass. These quantities are approximately linear in $C_k$ (see Fig.~\ref{Ck_Cn_fix}(a)), which allows to approximate the oscillation of the center of mass using a mixed modulation \cite{Ozimek1987}
\begin{eqnarray}
  y_{cm}^\text{MM}(t) & = & y^0_{cm}  [1-m \cos(2\pi f_{m} t)] \nonumber \\
  &&\; \;\;\;\;\;\;\; \times\cos{[2\pi f_0 t + \beta \sin(2\pi f_{m} t)]}
  \; ,
  \label{eq:mixedmodulation}
\end{eqnarray}
where $m=\frac{0.13 \times C_k^0}{y^0_{cm}}\varepsilon_{m}$ and $\beta = \frac{f_0}{f_m}\varepsilon_{m}$ are the indices of, respectively, amplitude and frequency modulation (see App.~\ref{app:analyticalmodel} for details) \footnote{Note that we do not take into account the higher harmonics of the initial CSn in this approach to simplify the discussion.}. The envelope of $y_{cm}^\text{MM}$ oscillates between $y^+=y^0_{cm} (1 + m)$ and $y^-=y^0_{cm} (1 - m)$ and does not depend on the frequency $f_m$ of the flow.  

One can easily show that:
\begin{eqnarray}
  \frac{y_{cm}^\text{MM}(t)}{y^0_{cm}} & = & \cos{(2\pi f_0 t)} - \frac{(\beta+m)}{2}  \cos{[2\pi (f_0 - f_m) t]}  \nonumber\\ 
  &&\;\;\; 
  + \frac{(\beta - m)}{2}  \cos{[2\pi (f_0 + f_m) t]} + \ldots
  \; ,
\end{eqnarray}
to lowest order in $m$ and $\beta$. The resulting spectrum consists of the fundamental frequency $f_0$ of the initial CSn and sidebands with frequencies $f_{0}\pm n f_{m}$ $(n\in\NN)$, which are due to the mixed modulation. 
We focus on the point B in the following and fix $\varepsilon_{m}=0.1$. 
Fig.~\ref{MM} shows for $f_{m}=0.01$ that the analytical approximation, Eq.~(\ref{eq:mixedmodulation}), describes the result of the simulation surprisingly well. Even the amplitudes of the sidebands around the fundamental frequency $f_0$ (see Fig.~\ref{MM}(d)) are predicted correctly with our simple model.

\begin{figure}
 \includegraphics[width=\columnwidth]{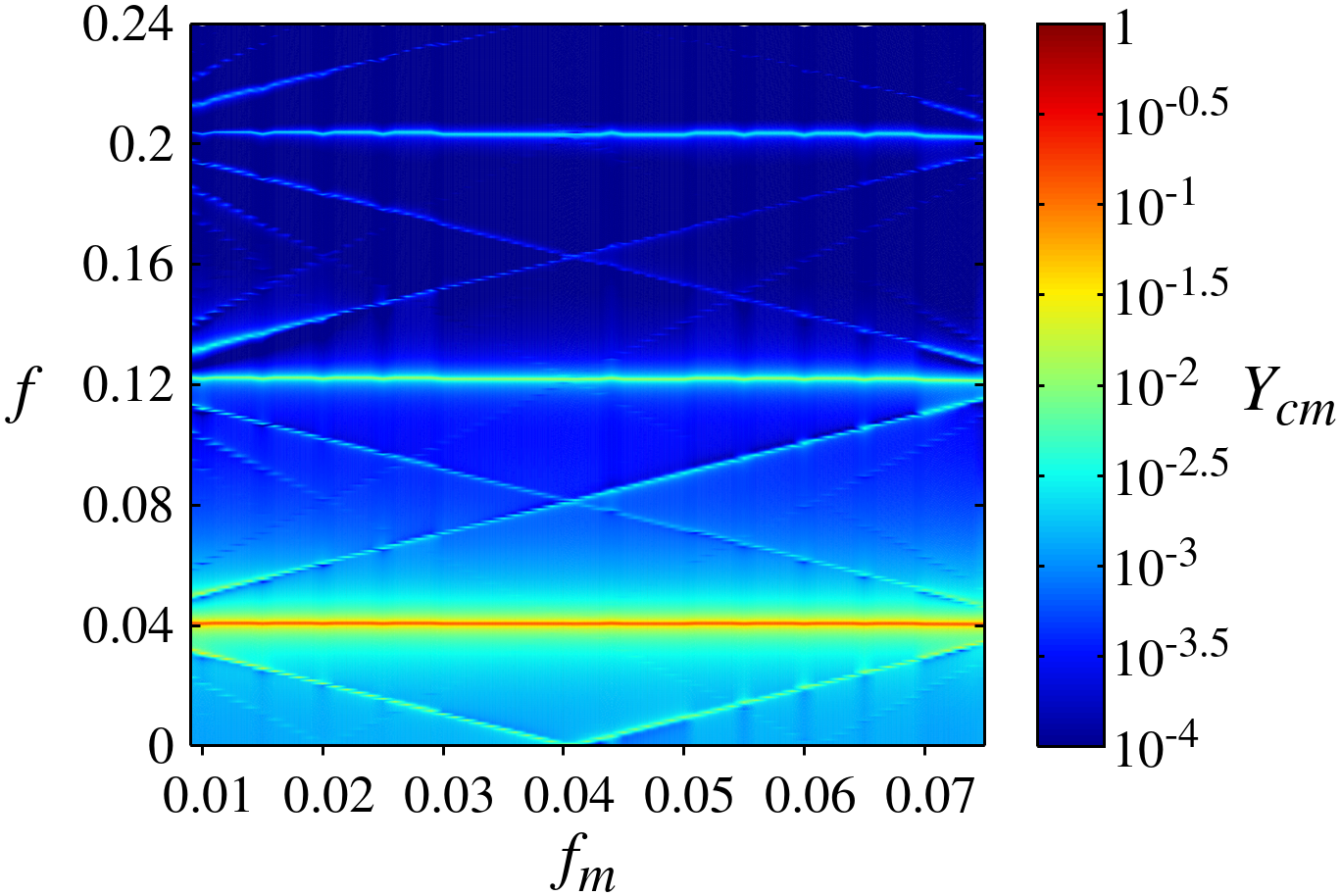}
 \caption{Fourier transform of $y_{cm}$ from the simulation for a fixed modulation amplitude $\varepsilon_{m}=0.1$ and modulation frequencies $f_{m}$ ranging from $0.009$ to $0.08$. The frequency $f$ of the Fourier transform is depicted on the vertical axis and the amplitude $Y_{cm}$ is shown using a heat map. $Y_{cm}$ is scaled by its maximum value whereas all other variables are scaled as described in Sec.~\ref{sec:themodel}.}\label{phase_fft}
\end{figure}

This observation could be confirmed for other modulation frequencies as well. Fig.~\ref{phase_fft} shows the Fourier transform of simulation results for a fixed perturbation amplitude $\varepsilon_{m}=0.1$ and varying $f_{m}$. 
The first frequencies of the sidebands are clearly visible and coincide with what is predicted by the analytical theory, not only for the fundamental frequency of the initial CSn shape $(f_0=0.04)$ but also for the higher harmonics $3 f_0$ and $5 f_0$.

\subsubsection{\emph{Oscillations with constant envelope}} 

\begin{figure}[h!]
  \includegraphics[width=\columnwidth]{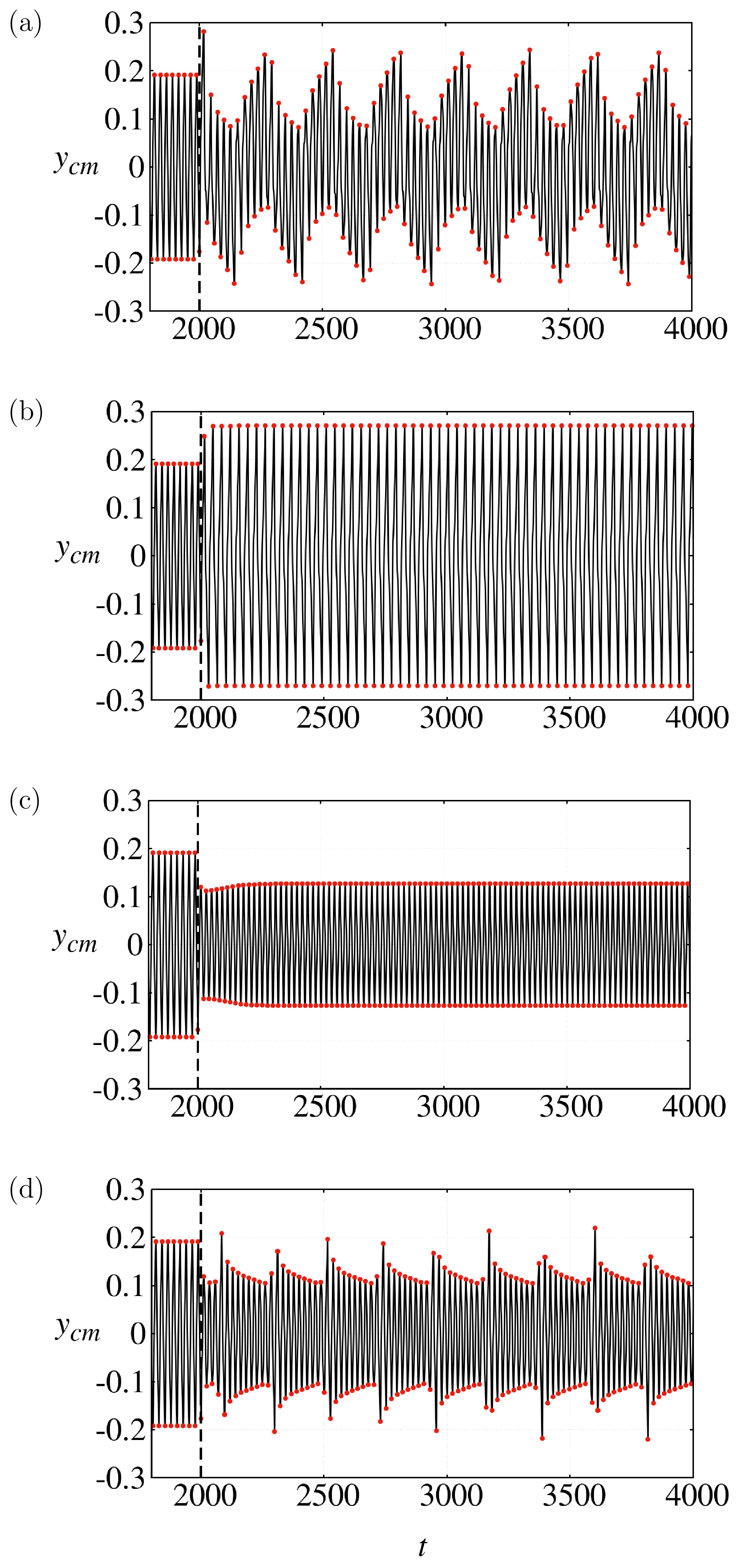}
  \caption{Vertical component of the vesicle's center of mass, $y_{cm}$, as a function of time for a fixed modulation amplitude 
  $\varepsilon_{m}=0.7$ and frequencies (a) $f_{m}=0.04$, (b) $f_{m}=0.056$, (c) $f_{m}=0.092$, (d) $f_{m}=0.093$. The dashed line represents the time at which the modulation of the Poiseuille flow is switched on. All variables are scaled as described in Sec.~\ref{sec:themodel}.}\label{oscillation_modes}
\end{figure}

When the modulation frequency $f_m$ of the flow is exactly a multiple of the fundamental frequency, $f_m=n f_0$  $(n\in\NN)$, the upper and lower envelopes of $y_{cm}$ should be constants according to the analytical theory, Eq.~(\ref{eq:mixedmodulation}). For an even $n$ the total envelope should be symmetric with respect to the center of the channel whereas it should be asymmetric for odd $n$. This implies that there are two solutions with constant envelope for odd $n$: one with envelope situated at $y^+$ and  $- y^-$ and a second one with envelope at  $y^-$ and  $- y^+$. 
To check these predictions we again focus on the point B in the following and fix $\varepsilon_{m}$ to a value smaller than one to avoid any transition to a static shape (see Sec.~\ref{subsubsec:amplitude}). Looking at the simulations, we observe that the oscillation of $y_{cm}$ \textit{does not} display a constant envelope for $f_m=f_0$. The numerical results indicate that the two solutions with \textit{asymmetric} constant envelope ($n$ is odd) are linearly unstable; the system oscillates between them. 
However, one indeed finds a constant envelope for $f_m = 2 f_0$, which is symmetric with respect to the channel. But even when $f_m$ is not exactly $2f_0$, the envelope of the signal equals a constant whose value depends on $\varepsilon_{m}$ and $f_m$. Fig.~\ref{oscillation_modes} shows one example for fixed $\varepsilon_{m}=0.7$. 
For $f_{m}=f_0=0.04$ the vesicle oscillates between the two asymmetric solutions of constant envelope as mentioned above (see Fig.~\ref{oscillation_modes}(a)).

Increasing the modulation frequency with a step size of $10^{-3}$, the vesicle exhibits stable oscillations already at $f_{m}=0.056 < 2 f_0$ 
(see Fig.~\ref{oscillation_modes}(b)). When increasing the frequency $f_{m}$ even more, the envelope of the oscillation remains a constant while attenuating until the frequency $f_{m}=0.092>2f_0$ is reached (Fig.~\ref{oscillation_modes}(c)). Above this frequency the envelope of $y_{cm}$ is not a constant again. 

\begin{figure}
  \includegraphics[width=\columnwidth]{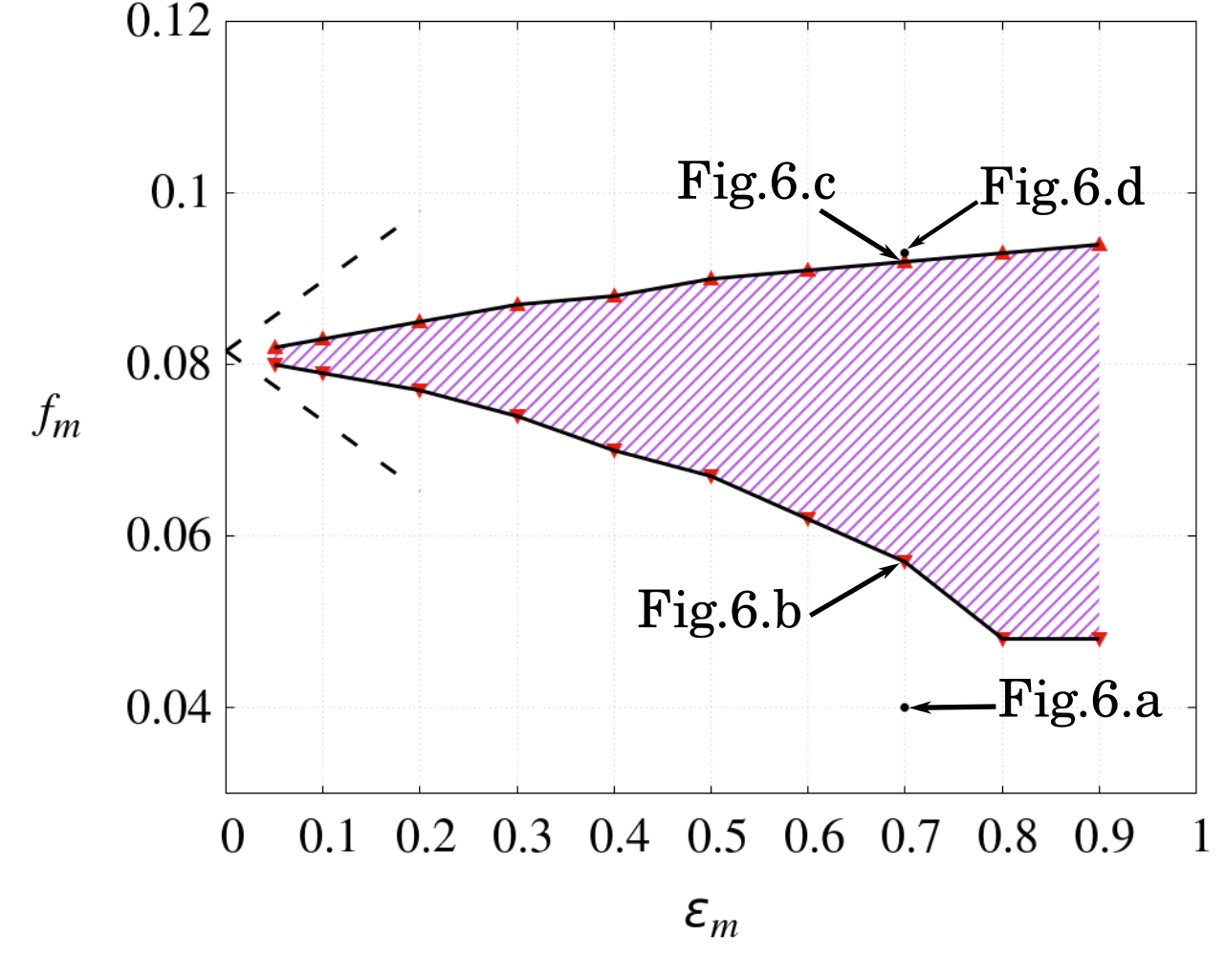}
  \caption{Region of oscillations with constant envelope (hashed) as a function of the imposed amplitude $\varepsilon_{m}$ and frequency $f_{m}$ of the harmonically modulated flow. The dashed lines indicate an analytical estimate of the boundaries of this region for small $\varepsilon_{m}$ (see text).}\label{phase_diag_freq}
\end{figure}

To find the domain of frequencies where the oscillation displays a constant envelope in the simulation, we have varied the modulation amplitude $\varepsilon_{m}$ from 0.1 to 0.9 with a step size of 0.1 \footnote{For $\varepsilon_{m}=1$ the shape of the vesicle will evolve to a parachute as discussed earlier.} and the imposed frequency $f_{m}$ with $10^{-3}$.
Fig.~\ref{phase_diag_freq} presents the resulting region (hashed area). Its size increases with increasing $\varepsilon_m$. 

How can this surprising behavior be explained?  
The answer lies in the observation that the imposed flow not only corresponds to an oscillation around the point B but can also be interpreted as an oscillation around \emph{other} points B' of same confinement $C_n=0.55$, as long as their ``stationary" capillary number $C_k^{0'}$ lies in the range of capillary numbers that are reached by the oscillation of the flow. The points B' all lie on the same vertical line of the phase diagram as point B (see dashed line in Fig.~\ref{phase_diag}). 

A closer look into the results of the simulation reveals that $y_{cm}$ oscillates with a fundamental frequency of $f_0'=f_m/2$ when its envelope is constant. This oscillation corresponds to a mixed modulation around another point of the phase diagram whose ``stationary'' CSn has the frequency $f_0'$. Consequently, there can be two different states depending on the values of $f_m$ and $\varepsilon_m$ in our system. State 1 can be understood as a mixed modulation of the ``stationary'' CSn of point B (see previous subsection), whereas state 2 is a mixed modulation around point B' with fundamental frequency $f_0'$ and a constant envelope. When $f_m$ is close to $2f_0$, the system is in state 2. However, for this state to be accessible, $f_0'$ has to lie in the range of steady state frequencies (compare Fig.~\ref{Ck_Cn_fix}(a)) of the points on the vertical line of the phase diagram that the flow reaches during its oscillation. If we take this criterion as an estimate for the boundaries of the region of oscillations with constant envelope, we obtain the dashed lines depicted in Fig.~\ref{phase_diag_freq}.

The discrepancy between this theoretical prediction and the simulations is most probably due to the fact that the oscillation in $C_k$ is only symmetric around the point B but asymmetric for all other points B'. One thus expects a transition region where the system switches between states 1 and 2.  This is indeed what we observe: there are frequencies (next to the boundaries of the region of constant envelope) where the system reaches an envelope of constant amplitude which then starts oscillating again. This behavior repeats itself periodically. Further complications arise because 
the linear approximation of amplitude and frequency, Eq.~(\ref{eq:amplitudefrequencyfits}), for the mixed modulation breaks down for modulation amplitudes larger than $\varepsilon_m\approx 0.2$. When $\varepsilon_m$ is close to one we even expect that there are further states that the vesicle might be able to adopt. Simulations for $\varepsilon_m=0.9$ indeed show that the vesicle tries to take on a shape resembling a confined slipper but does not quite manage to do so. The underlying theory goes beyond the scope of this paper but could be a promising subject of future studies.


\section{Conclusions}
In this paper we have studied the snaking oscillations of a membrane vesicle in a channel under two types of Poiseuille flow. 
In a steady Poiseuille flow the behavior of the vesicle only depends on the width of the channel and the ratio between flow strength and bending energy of the membrane. The resulting two-dimensional phase diagram consists of regions in which different shapes are found. The region of centered snaking shapes (CSn) falls into three different parts: two regions of coexistence where---depending on the initial conditions---either parachutes or unconfined slippers are found in addition to the CSn and a region at higher confinement where only the CSn is found. To characterize the oscillation of the CSn, we have studied the motion of the vesicle's center of mass as a function of time. An increase of either the flow strength or the degree of confinement forces the vesicle to oscillate faster but with decreasing amplitude. The oscillations attenuate until the vesicle makes a transition to either a parachute (large relative flow strength) or an off-centered snaking shape (small channel width).

Using a harmonically modulated Poiseuille flow we have then perturbed a CSn lying in the center of the CSn region of the phase diagram. When the modulation amplitude of the flow is large enough, transitions to static shapes are observed. For small modulation amplitudes the vesicle keeps oscillating but in a more complex manner than the CSn. The corresponding motion of the vesicle's center of mass is nicely described with a mixed modulation: an amplitude modulation of the flow induces a modulation in amplitude \textit{and} frequency of the vesicle's center of mass. The exact behavior of the vesicle depends on the modulation amplitude and the frequency of the flow. Surprisingly, we find that the center of mass oscillates with a constant envelope in a certain parameter range. This indicates that there are (at least) two stable states. A more detailed theoretical study of this phenomenon is certainly interesting for future work. The associated stabilization of the vesicle's oscillation could be of interest for technical applications as well. 


\begin{acknowledgments}
The authors thank the French-German University (DFH-UFA, ``Living Fluids'' DFDK/CFDA-01-14) and C.\ M. thanks the CNES (Centre National d'Etudes Spatiales) for financial support.
\end{acknowledgments}


\appendix

\section{Analytical model\label{app:analyticalmodel}}

In Sec.~\ref{sec:modulatedPoiseuille} a vesicle with a CSn in steady Poiseuille flow is put into a harmonically modulated Poiseuille flow, Eq.~(\ref{eq:pois}). This implies an oscillation of the capillary number $C_k$ of the system which is directly proportional to the imposed flow (see Sec.~\ref{subsubsec:frequency}). $C_k$ is thus harmonically modulated with the same amplitude $\varepsilon_m$ and frequency $f_m$ as the flow:
\begin{equation}
  C_k(t) = C_k^0 [1 + \varepsilon_{m} \cos(2\pi f_{m} t) ]
  \; ,
  \label{eq:Ck_t}
\end{equation}
where $C_k^0$ is the capillary number corresponding to the initial CSn. In other words, we force the system to oscillate on a vertical line of the phase diagram (Fig.~\ref{phase_diag}).

From the data of Fig.~\ref{Ck_Cn_fix}(a) we can obtain linear fits for amplitude and fundamental frequency of the vertical component of the vesicle's center of mass as a function of $C_k$:
\begin{equation}
  y^\text{max}_{cm}(C_k) = y^0_{cm} - a (C_k - C_k^0), \quad \text{and} \quad f(C_k) = \frac{f_0}{C_k^0} \, C_k
  \; ,
  \label{eq:amplitudefrequencyfits}
\end{equation}
with $a=0.13$. The value of $a$ is obtained by neglecting the points for which $C_k<0.3$. The fit of $f$ is only acceptable for $C_k<1.6$ since the error becomes too large for higher values of $C_k$. The parameters $y^0_{cm}$ and $f_0$ are the fundamental frequency and capillary number of the initial CSn. 

Inserting $C_k(t)$ from Eq.~(\ref{eq:Ck_t}) into these fits yields
\begin{equation}
  y^\text{max}_{cm}(t) = y^0_{cm} \left[1 - m \cos(2\pi f_{m} t) \right]
\end{equation}
with $m=\frac{a C_k^0}{y^0_{cm}}\varepsilon_{m}$ 
for the amplitude and 
\begin{equation}
 f(t) = f_0 [1 + \varepsilon_{m} \cos(2\pi f_{m} t) ]
\end{equation}
for the frequency as a function of time. The whole signal thus consists of a simultaneous amplitude and 
frequency modulation. This mixed modulation (MM) can be written as
\begin{equation}
   y_{cm}^\text{MM}(t) =  y^\text{max}_{cm}(t) \cos{\theta(t)}
\end{equation}
with phase angle
\begin{equation}
 \theta(t) = 2\pi \int_0^t f(t')\text{d}t' = 2\pi f_0 t + \beta \sin(2\pi f_{m} t)
 \; ,
\end{equation}
where $\beta =\frac{f_0}{f_m}\varepsilon_{m}$ is the index of the frequency modulation. 
Putting everything together we obtain Eq.~(\ref{eq:mixedmodulation}):
\begin{eqnarray}
  y_{cm}^\text{MM}(t) & = & y^0_{cm}  [1-m \cos(2\pi f_{m} t)] \nonumber \\
  &&\; \;\;\;\;\;\;\; \times\cos{[2\pi f_0 t + \beta \sin(2\pi f_{m} t)]}
  \; .
\end{eqnarray}

\bibliography{biblio} 

\clearpage

\end{document}